\def\year{2020}\relax
\documentclass[letterpaper]{article} 
\usepackage{aaai20}  
\usepackage{times}  
\usepackage{helvet} 
\usepackage{courier}  
\usepackage[hyphens]{url}  
\usepackage{graphicx} 
\usepackage{graphicx}  
\usepackage[utf8]{inputenc} 
\usepackage{url}            
\usepackage{booktabs}       
\usepackage{amsfonts}       
\usepackage{nicefrac}       
\usepackage{microtype}      
\usepackage{graphicx}
\usepackage{makecell}
\usepackage[ruled,vlined,linesnumbered]{algorithm2e}
\usepackage[caption=false]{subfig}
\usepackage{amsmath,amsthm,amssymb}
\usepackage{cleveref}
\usepackage{bibunits}

\newcommand{\citet}[1]{\citeauthor{#1}~\shortcite{#1}}

\crefformat{footnote}{#2\footnotemark[#1]#3}

\newtheorem{theorem}{Theorem}
\newtheorem{corollary}{Corollary}

\newtheorem{claim}{Claim}
\DontPrintSemicolon

\title{Draining the Water Hole:\\ Mitigating Social Engineering Attacks with CyberTWEAK}

\author{Zheyuan Ryan Shi,\textsuperscript{\rm 1} Aaron Schlenker,\textsuperscript{\rm 2} Brian Hay\textsuperscript{\rm 3}\\ \Large \textbf{Daniel Bittleston,\textsuperscript{\rm 1} Siyu Gao,\textsuperscript{\rm 1} Emily Peterson,\textsuperscript{\rm 1} John Trezza,\textsuperscript{\rm 1} Fei Fang\textsuperscript{\rm 1}}\\ 
	\textsuperscript{\rm 1}Carnegie Mellon University, \textsuperscript{\rm 2}Facebook, Inc., \textsuperscript{\rm 3}Security Works\\ 
	ryanshi@cmu.edu, aschlenker@fb.com, bhay@securityworks.com\\ \{dbittles, siyug, emilypet, jtrezza\}@andrew.cmu.edu, feif@cs.cmu.edu
}

\begin{document}
\maketitle
\begin{abstract}
Cyber adversaries have increasingly leveraged social engineering attacks to breach large organizations and threaten the well-being of today's online users. One clever technique, the ``watering hole'' attack, compromises a legitimate website to execute drive-by download attacks by redirecting users to another malicious domain.  
We introduce a game-theoretic model that captures the salient aspects for an organization protecting itself from a watering hole attack by altering the environment information in web traffic so as to deceive the attackers.
Our main contributions are (1) a novel Social Engineering Deception (SED) game model that features a continuous action set for the attacker, 
(2) an in-depth analysis of the SED model to identify computationally feasible real-world cases, and
(3) the \textsc{CyberTWEAK} algorithm which solves for the optimal protection policy.
To illustrate the potential use of our framework, we built a browser extension based on our algorithms which is now publicly available online. The \textsc{CyberTWEAK} extension will be vital to the continued development and deployment of countermeasures for social engineering.
\end{abstract}

\section{Introduction}

Social engineering attacks are a scourge for the well-being of today's online user and the current threat landscape only continues to become more dangerous~\cite{mitnick2001art}. Social engineering attacks manipulate people to give up confidential information through the use of phishing campaigns, spear phishing whaling or watering hole attacks.
For example, in watering hole attacks, the attacker compromises a legitimate website and redirects visitors to a malicious domain where the attacker can intrude the user's network.
The number of social engineering attacks is growing at a catastrophic rate.
In a recent survey, 60\% organizations were or may have been victim of at least one attack~\cite{ciosummits}. 
Such cybercrime poses an enormous threat to the security at all levels -- national, business, and individual.

To mitigate these attacks, organizations take countermeasures from employee awareness training to technology-based defenses. 
Unfortunately, existing defenses are inadequate. Watering hole attackers typically use zero-day exploits, rendering patching and updating almost useless~\cite{sutton2014}. 
Sand-boxing potential attacks by VM requires high-end hardware, which hinders its wide adoption~\cite{farquhar2017}. White/blacklisting websites is of limited use, since the adversary is strategically infecting trustworthy websites. 

We propose a game-theoretic deception framework to mitigate social engineering attacks, and, in particular, the watering hole attacks.
Deception is to delay and misdirect an adversary by incorporating ambiguity. 
Watering hole attackers rely on the identification of a visitor's system environment to deliver the correct malware to compromise a victim. 
Towards this end, the defender can manipulate the identifying information in the network packets, such as the user-agent string, IP address, and time-to-live.
Consequently, the attacker might receive false or confusing information about the environment and send incompatible
exploits.
Thus, deceptively manipulating employees' network packets provides a promising countermeasure to social engineering attacks.

\textbf{Our Contributions \,\,} 
We provide the first game-theoretic framework for autonomous countermeasures to social engineering attacks.
We propose the Social Engineering Deception (SED) game,
in which an organization (defender) strategically alters its network packets. 
The attacker selects websites to compromise, and captures the organization's traffic to launch an attack.
We model it as a zero-sum game and consider the minimax strategy 
for the defender.

Second, we analyze the structure and properties of the SED game, based on which we identify real-world scenarios where the optimal protection policy can be found efficiently.

Third, we propose the \textsc{CyberTWEAK} (Thwart WatEring hole AttacK) algorithm to solve the SED game. \textsc{CyberTWEAK} exploits theoretical properties of SED, linear program relaxation of the attacker's best response problem, and the column generation method, and is enhanced with dominated website elimination.
We show that our algorithm can handle corporate-scale instances involving over $10^5$ websites.

Finally, we have developed a browser extension based on our algorithm. The software is now publicly available on the Chrome Web Store.\footnote{\label{fnt:extension}\url{http://bit.ly/CyberTWEAK}} The extension is able to manipulate the user-agent string in the network packets. We take additional steps to improve the its usability and explain the output of \textsc{CyberTWEAK} intuitively. We believe it will be vital to the continued development of social engineering defenses.

\textbf{Related Work\,\,}
Deception is one of the most effective ways to thwart cyberattacks.
Recent papers have considered deception techniques for protecting an enterprise network from an attack by sending altered system environment information in response to scans performed during the reconnaissance phase of an attack~\cite{albanese2016deceiving,jajodia2017probabilistic}.
There is a rising interest in building game-theoretic models for deception~\cite{schlenker2018deceiving}, in particular in the use of honeypots~\cite{durkota2015optimal,pibil2012game} in the enterprise network. 

However, there is a fundamental difference between enterprise network defense and social engineering defense. In the former, an adversary targets an organization by compromising computers in the network while in watering hole attacks the attacker targets the user and compromises external websites. 
A website in SED cannot be properly modeled as a honeypot target, because the defender has no control over it. Neither can the user, because the attack depends on an external task -- compromising a website.
Instead of actively querying the network, watering hole attackers passively monitor the users' traffic. This necessitates the continuous action space for the attacker in SED, 
which is also different from most previous works on enterprise network defense.

\citet{laszka2015optimal} study spear phishing, another form of social engineering attacks. The nature of watering hole attacks leads to additional complications. For example, watering hole attackers need to compromise a website and then scan the traffic. 
Thus, in SED the attacker has two layers of decision making: one continuous and one discrete. This leads to a different problem formulation and solution techniques than those in spear phishing.

\section{Watering Hole Attacks}

\begin{figure}[t]
\begin{center}
  \includegraphics[width=0.9\columnwidth]{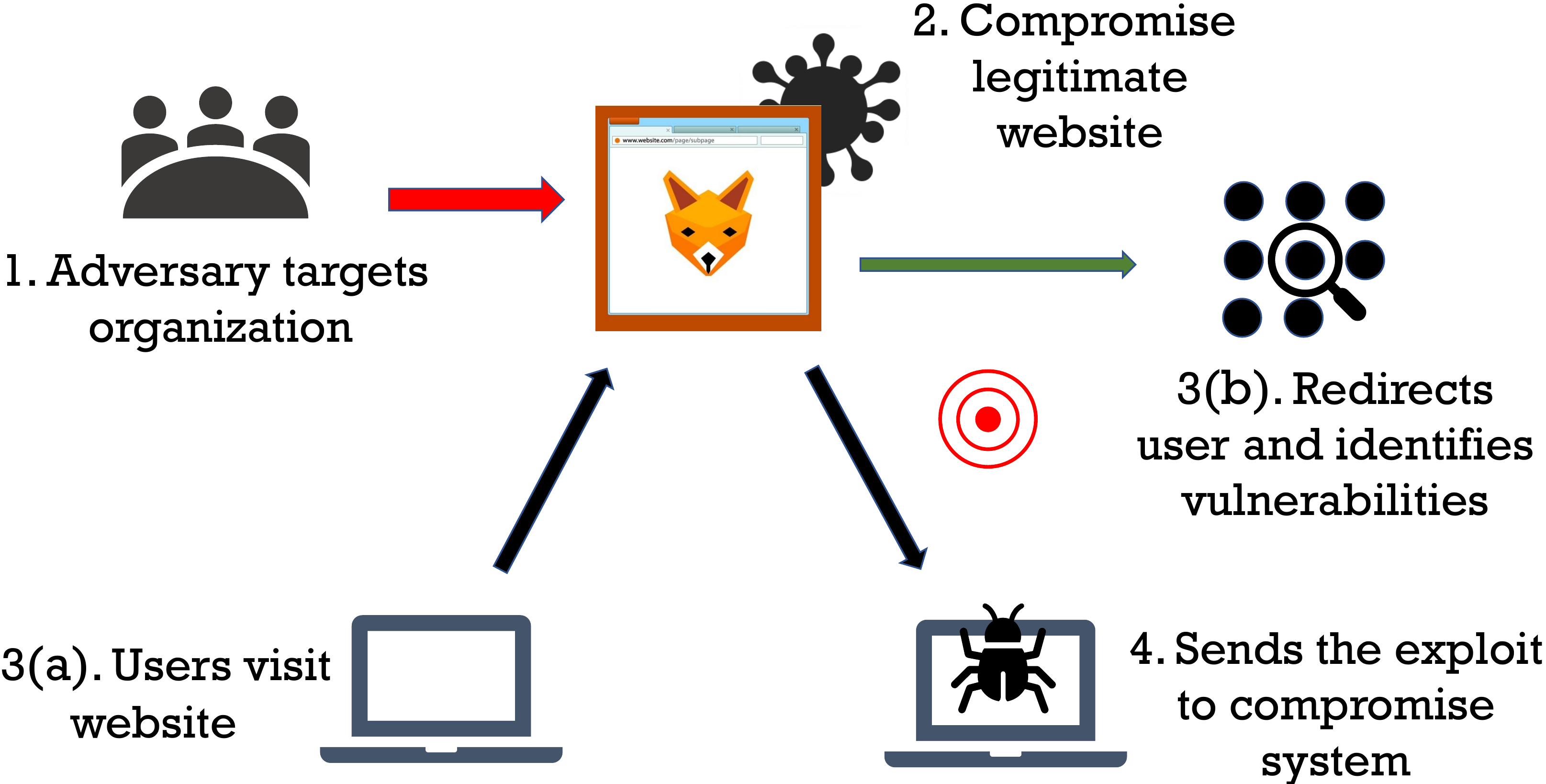} 
  \caption{Anatomy of a watering hole attack. }
  \label{fig:domain}
  \end{center}
\end{figure}

Watering hole attacks are a prominent type of social engineering used by sophisticated attackers. 
Before we describe our modeling decisions, it is useful to highlight the primary steps in executing a watering hole attack, as illustrated in Fig.~\ref{fig:domain}.
In step 1, the attacker identifies a target organization. They use surveys and external information like specialized technical sites to understand the browsing habits of its employees.
This allows the adversary to determine the most lucrative websites to compromise for maximum exposure to employees from the targeted organization.
In step 2, the adversary compromises a set of legitimate websites. Not only do these websites need to be lucrative, but the attacker also has to be strategic in this choice. For example, compromising Google.com is nearly impossible while the Polish Financial Authority, victim of the 2017 Ratankba malware attacks~\cite{symantec2017}, cannot invest the same security resources.  
Indeed, in previous attacks the attacker was not observed to compromise all websites~\cite{cyberpro2018}.
In step 3, employees visit the compromised website and are redirected to a malicious website which scans their system environment and the present vulnerabilities.  
To gather this information, attackers use techniques such as analyzing the user-agent string, operating system fingerprinting, etc.
In Step 4, the attacker delivers an exploit for an identified vulnerability. After these steps, the attacker can navigate the target network and access the sensitive information. 

Our algorithm and browser extension introduce uncertainty in step 3 of a watering hole attack.  
Identifying the vulnerabilities in a visitor relies on the information gathered from reconnaissance.  
The extension modifies the network packets so that the attacker gets false information about the visitor.
Deception is not free, though. Altering the network packet can degrade the webpage rendered, e.g., displaying for Android on a Windows desktop. Thus, the defender needs to carefully trade-off security and the quality of service.

In reality, sophisticated attackers typically do not send all exploits without tailoring to the packet information, as defense would become easier after seeing more such unknown exploits. Also, sending all exploits would be flagged as suspicious and get blocked. The attacker would need to get a new zero-day -- a costly proposition. 
Thus, the attacker prefers scanning the system environment of the incoming traffic.

\section{Social Engineering Deception Game} \label{sec:model}
We model the strategic interaction between the organization (defender) and an adversary as a two-player zero-sum game,
where the defender chooses an alteration policy
and the adversary chooses which websites to compromise and decides the effort spent on scanning traffic.  
In everyday activities employees of a target organization $O$
visit a set of websites $W$ which includes legitimate sites and potential watering holes set up by an adversary. Let $t^{all}_w$ denote the total amount of traffic to $w\in W$ from all visitors and $t_w$ the total traffic to $w$ from $O$.
The defender's alteration policy is represented by $x\in [0,1]^{|W|}$ where $x_w$ is the proportion of $O$'s traffic to website $w \in W$ for which the network packet will be altered. 
We assume a drive-by download attack will be unsuccessful if, and only if an employee's packet is altered. 
However, it is easy to account for different levels of adversary and defender sophistication by adding an additional factor in Eq.~\eqref{lp1Eq1} below.
We consider a cost $c_w$ to alter a single unit of traffic to $w$. 
The defender is limited to a budget $B_d$ on the allowable cost.

The adversary first chooses which websites to compromise, represented by a binary vector $y \in \{0,1\}^{|W|}$. If $y_w=1$, i.e., they turn website $w$ into a watering hole, they must pay a cost $\pi_w$. The attacker has a budget $B_a$ for compromising websites (w.l.o.g. we assume $\pi_w \leq B_a\ \forall w\in W$).
The adversary then decides the scanning effort for each compromised website which can enable them to send exploits tailored to the packet information. We use $e_w$ to denote how much traffic the attacker decides to scan per week for $w$, and refer to $e$ as the \emph{effort vector}.
The discreet attacker has a budget $B_e$ for scanning the incoming traffic.
In the special case where the scanning effort is negligible ($B_e = \infty$), all our complexity and algorithmic results to be introduced still hold.

We consider an attacker who aims to maximize the expected amount of unaltered flow from target organization $O$ that is scanned by them, as each unit of scanned unaltered flow can lead to a potential success in the social engineering attack, i.e., compromise an employee and discover critical information about $O$. We model it as a zero-sum game, and therefore the defender's goal is to minimize this amount.

Social engineering is a complex domain which we cannot fully model. However, we build our model and assumptions so that we can formally reason about deception, and even when our assumptions are not met, our work provides a sensible solution. For example,
cyber attackers may have tools to circumvent existing deception techniques. 
Nonetheless, our solution increases the attacker's uncertainty about the environment as they cannot easily obtain or trust the information in the network packets.
In Appendix~\ref{app:discussion}, we provide a detailed discussion of the generality and limitations of our work.

\section{Computing Optimal Defender Strategy}
In this section, we present complexity analysis and algorithms for finding the optimal defender strategy $x^*$ in this game, which is essentially the minimax strategy, i.e., a strategy that minmizes the attacker's maximum possible expected amount of scanned unaltered flow. 
$x^*$ should be the solution of the following bi-level optimization problem $\mathcal{P}_1$.
\begin{align}
\mathcal P_1: \,\min\nolimits_{x}&  \max\nolimits_{y,e} \quad  \sum\nolimits_{w\in W} \kappa_w(1-x_w)e_w \label{lp1Eq1}\\
\text{s.t.}  &  \quad \sum\nolimits_{w\in W} e_w\leq B_e   \label{lp1Eq2}\\
& \quad \sum\nolimits_{w\in W} \pi_w y_w \leq B_a  \label{lp1Eq3}\\
 & \quad  e_w\leq t^{all}_w \cdot y_w, \forall w\in W \label{lp1Eq4}\\
 &\quad y_w\in \{0,1\}, \forall w\in W \label{lp1Eq5+}\\
 &\quad e_w \in [0, \infty), \forall w\in W  \label{lp1Eq5}\\
 & \quad \sum\nolimits_{w\in W} c_w t_w x_w\leq B_d  \label{lp1Eq6}\\ 
&\quad x_w\in [0,1], \forall w\in W  \label{lp1Eq7}
\end{align}
In objective function \ref{lp1Eq1}, $\kappa_w=t_w/t_w^{all}$. Since $t_w(1-x_w)$ is the total amount of unaltered flow from the defender organization $O$ and $e_w/t_w^{all}$ is the percentage of incoming traffic that will be scanned, $\kappa_w(1-x_w)e_w$ is the total scanned unaltered traffic to $w$. Constraint \ref{lp1Eq2}-\ref{lp1Eq3} describes the budget constraint for the attacker, and Constraint \ref{lp1Eq4} requires that the attacker can only scan traffic for the compromised websites. Constraint \ref{lp1Eq6} is the budget constraint for the defender.

Unfortunately, solving $\mathcal{P}_1$ is challenging. It cannot be solved using any of the existing solvers directly due to the bi-level optimization structure, the mix of real-valued and binary variables and the bilinear terms in the objective function ($x_we_w$). In fact, even the adversary's best response problem $\mathcal P_2(x)$, represented as a mixed integer linear program (MILP) below, is NP-hard as stated in Thm \ref{thm:abr}. Due to space limit, we defer all the proofs to appendix.\footnote{\url{https://arxiv.org/abs/1901.00586}}
\begin{align}
\mathcal P_2(x): \quad \max_{y,e} & \quad  \sum\nolimits_{w\in W} \kappa_w(1-x_w)e_w & \label{lp2Eq1}\\
\ \ \  \text{s.t.}  &  \quad \text{Constraints } \eqref{lp1Eq2}\sim\eqref{lp1Eq5}  & \label{lp2Eq2} 
\end{align}
\begin{theorem}\label{thm:abr}
Finding adversary's best response is NP-hard.
\end{theorem}

Therefore, we exploit the structure and properties of SED and $\mathcal{P}_1$ and design several novel algorithms to solve it. We first identify two tractable special classes of SED games which can be solved in polynomial time and discuss their real world implications. Then we present \textsc{CyberTWEAK}, our algorithm for general SED games.

\subsection{Tractable Classes} The first tractable class is identified based on the key observation stated in Thm \ref{thm:greedy-allocation}: the optimal solutions of SED games exhibit a greedy allocation of the attacker's effort budget. That is, for at most one website $w$ will the attacker spend scanning effort neither zero nor $t_w^{all}$. 
\begin{theorem} \label{thm:greedy-allocation}
Let $(x^*, y^*, e^*)$ be an optimal solution to $\mathcal P_1$, $W_F = \{w: e^*_w = t^{all}_w\}, W_Z = \{w: e^*_w = 0\}, W_B = \{w: e^*_w \in (0, t^{all}_w)\}$. There is an optimal solution with $|W_B| \leq 1$.
\end{theorem}
As a result, if the attacker's scanning budget is so limited that he cannot even scan through the traffic of any website, he will use all the scanning effort on one website in the optimal solution.
Thus, the optimal defender strategy can be found by enumerating the websites.
\begin{corollary} \label{cor:smallbudget}
	(Small Effort Budget) If $0 < B_e \leq t^{all}_w, \forall w$, the optimal solution can be found in polynomial time.
\end{corollary}

The second tractable class roots in the fact that if the scanning effort is negligible (or equivalently, $B_e=\infty$) the attacker only needs to reason about which websites to compromise. Further, if the attacker has a systematic way of compromising a website which makes the cost $\pi_w$ uniform across websites, then the attacker only needs to greedily choose the websites with the highest unaltered incoming traffic and the defender can greedily alter traffic in the top websites. We provide details about these algorithms in the appendix.
\begin{theorem} \label{thm:unifcost}
(Uniform Cost $+$ Unlimited Effort) If $\pi_w = 1, \forall w\in W$ and $B_e = \infty$, the defender's optimal strategy can be found in polynomial time.
\end{theorem}

\subsection{CyberTWEAK}
For the general SED games, we propose a novel algorithm \textsc{CyberTWEAK} (Alg \ref{alg:cutgen}). It first computes an upper bound for $\mathcal{P}_1$ leveraging the dual problem of the linear program (LP) relaxation of $\mathcal P_2(x)$. As a byproduct, the computation provides a heuristic defender strategy $\hat{x}^*$ (Line \ref{algline:relaxedlp}). It then runs an optimality check (Line \ref{algline:checkopt}) to see if $\hat{x}^*$ is optimal for $\mathcal P_1$.
When optimality cannot be verified, it solves the original problem $\mathcal P_1$ by converting $\mathcal P_1$ to an equivalent LP and applying column generation~\cite{gilmore1961linear}, an iterative approach to compute the optimal strategy (Line \ref{algline:cg_s}-\ref{algline:cg_e}).
We further improve the scalability by identifying and eliminating dominated website as pre-processing (Line \ref{algline:dwe}). Next we provide details about these steps.
\begin{algorithm}[t]
	Remove $D\leftarrow$\textsc{Find-dominated-websites()} from $W$.\label{algline:dwe}\\
	Get heuristic defender strategy $\hat{x}^*$ by solving $\hat{\mathcal P_1}$. \label{algline:relaxedlp}\\
	\lIf{$OPT(\mathcal P_2(\hat{x}^*))\leq OPT(\tilde{\mathcal P}_3(\hat{x}^*))$}{\Return $\hat{x}^*$ \label{algline:checkopt}}
	Initialize max effort vector set $e^{\mathcal A} = e^{\mathcal P_2(\hat{x}^*)}$.\label{algline:ini}\\
	\While{new max effort vector was added to $e^{\mathcal A}$ \label{algline:cg_s}}{
	    $x \gets$ solution of $P_1^{\text{LP}}(e^{\mathcal{A}})$.\label{algline:solvelp}\\
		$e \gets$ solution of $\mathcal P_2(x)$.\label{algline:attall}\\
		Add $e$ to $e^{\mathcal A}$.\label{algline:cg_e}
	} 
	\caption{\textsc{CyberTWEAK}}\label{alg:cutgen}
\end{algorithm}

\textbf{Upper Bound for $\mathcal P_1$\,\,\quad} 
Let $\hat{\mathcal P_2}(x)$ be the LP relaxation of $\mathcal P_2(x)$ and denote the dual variables of the (relaxed) constraints $\eqref{lp1Eq2}\sim\eqref{lp1Eq5+}$ as $\lambda_1,\lambda_2, \nu, \eta$.
We then include the variable $x$ for the defender strategy along with the dual problem, and obtain the minimization problem $\hat{\mathcal P_1}$.
\begin{align}
\hat{\mathcal P_1}: \min_{x, \lambda, \nu, \eta} & B_e\lambda_1 + B_a\lambda_2 + \sum\nolimits_{w \in W} \eta_w  \label{lp5Eq1}\\
\text{s.t.} \, & \,\kappa_w(1 - x_w) \leq \lambda_1 + \nu_w, \,\quad \forall w\in W  \label{lp5Eq2} \\
&\, \pi_w \lambda_2 - t^{all}_w \nu_w + \eta_{w} \geq 0, \qquad\forall w\in W    \label{lp5Eq3} \\
&\, \sum\nolimits_{w \in W} c_w t_w x_w \leq B_d & \label{lp5Eq4} \\
& x_w\in[0,1],\, \lambda_1, \lambda_2, \nu_w, \eta_{w} \geq 0, \,\forall w\in W  \label{lp5Eq5}
\end{align}
$\hat{\mathcal P_1}$ is an LP  which can be solved efficiently. In addition, $\hat{x}^*$ in the optimal solution for $\hat{\mathcal P_1}$ is a feasible defender strategy in the original problem $\mathcal P_1$. Therefore, solving $\hat{\mathcal P_1}$ leads to a heuristic defender strategy as well as bounds for the optimal value of $\mathcal P_1$. Denote the optimal value of a problem $\mathcal P$ as $\text{OPT}(\mathcal P)$. We formalize the bounds below.
\begin{theorem} \label{thm:3approx}
If $B_e \geq \max_w t_w^{all}$, $OPT(\hat{\mathcal P_1}) \leq 3 OPT(\mathcal P_1)$.
\end{theorem}
\begin{theorem} \label{thm:opt}
Let $x^*$, $\hat{x}^*$ be an optimal solution to $\mathcal P_1$, $\hat{\mathcal P_1}$. 
\begin{equation*} 
  \text{OPT}(\mathcal P_1) \leq \text{OPT}(\mathcal P_2(\hat{x}^*)) \leq \text{OPT}(\hat{\mathcal P_1}) \leq \text{OPT}(\hat{\mathcal P_2}(x^*)) . 
\end{equation*}
\end{theorem}

\textbf{Optimality Conditions for $\hat{x}^*$\,\,\quad} 
We present a sufficient condition for optimality, which leverages the solution of the following LP $\tilde{\mathcal P}_3(\hat{x}^*)$.
\begin{align}
\hspace{-0.5cm}& \tilde{\mathcal P}_3(\hat{x}^*): \min_{x, v} \quad v  \label{lp8Eq1}\\
\ \ \  \text{s.t.} & \quad v \geq \sum\nolimits_{w \in W} \kappa_w(1- x_w)e_w, \, \forall e\in e^{\mathcal P_2(\hat{x}^*)} \label{lp8Eq2}\\
& \quad \sum\nolimits_{w \in W} |x_w - \hat{x}^*| \leq \epsilon  \label{lp8Eq3} \\
& \quad \text{Constraints } \eqref{lp1Eq6}\sim\eqref{lp1Eq7} \nonumber
\end{align}
$\epsilon$ is an arbitrary positive number and $e^{\mathcal P_2(\hat{x}^*)}$ denotes the set of optimal effort vectors in $\mathcal P_2(\hat{x}^*)$. The following claim shows the optimality condition.
\begin{claim}\label{clm:lpopt}
Given $\hat{x}^*$, an optimal solution to $\hat{\mathcal P_1}$, $\hat{x}^*$ is optimal for $\mathcal P_1$ if $OPT(\mathcal P_2(\hat{x}^*))\leq OPT(\tilde{\mathcal P}_3(\hat{x}^*))$.
\end{claim}
Clearly, when $\epsilon$ is large, $OPT(\tilde{\mathcal P}_3(\hat{x}^*))$ is lower and it is harder to satisfy the condition, so in \textsc{CyberTWEAK}, we use a small enough $\epsilon$ in $\tilde{\mathcal P}_3(\hat{x}^*)$.

\textbf{Column Generation\,\,\quad} 
Define $\hat e^{\mathcal A}$ as the set of all \emph{max effort vectors} which satisfy $\sum_w e_w = B_e$ and $|W_B| \leq 1$. According to Thm~\ref{thm:greedy-allocation}, restricting the attacker to only choose strategies from $\hat e^{\mathcal A}$ will not impact the optimal solution for the defender.
As a result, $\mathcal P_1$ is equivalent to the following LP, denoted as $\mathcal P_1^{\text{LP}}(e^{\mathcal{A}})$, when $e^{\mathcal{A}}=\hat e^{\mathcal A}$.
\begin{align}
\hspace{-0.5cm} &
\mathcal P_1^{\text{LP}}(e^{\mathcal{A}}): \min_{x, v} \quad v \label{lp7Eq1}\\
\text{s.t.} & \quad v \geq \sum\nolimits_{w \in W} \kappa_w(1- x_w)e_w & \forall e\in e^{\mathcal{A}}  \label{lp7Eq2} \\
& \quad \text{Constraints } \eqref{lp1Eq6}\sim\eqref{lp1Eq7} \nonumber
\end{align}
Although existing LP solvers can solve $\mathcal P_1^{\text{LP}}(\hat e^{\mathcal A})$, the order of $\hat e^{\mathcal A}$ is prohibitively high, leading to poor scalability. 
Therefore, \textsc{CyberTWEAK} instead uses an iterative algorithm based on the column generation framework to incrementally generate constraints of the LP.
Instead of enumerating all of $\hat e^{\mathcal A}$, we keep a running subset $e^{\mathcal A} \subseteq \hat e^{\mathcal A}$ of max effort vectors and alternate between solving $\mathcal P_1^{\text{LP}}(e^{\mathcal A})$ (referred to as the master problem) and finding a new max effort vector to be added to $e^{\mathcal A}$ (slave problem).
In the slave problem, we solve the adversary's best response problem $\mathcal P_2(x)$ where $x$ is the latest defender strategy found.
This process repeats until no new effort vectors are found for the adversary. 
Recall that we get $\hat{x}^*$ and $e^{\mathcal P_2(\hat{x}^*)}$ when finding upper bound and verifying optimality of $\hat{x}^*$, which can serve as the initial set of strategies for column generation.

\textbf{Dominated Websites\,\,\quad} 
Not all websites are equally valuable for an organization as some are especially lucrative for an adversary to target. In a Polish bank, many employees may visit the Polish Financial Authority website daily, while perhaps a CS conference website is rarely visited by a banker. Intuitively, attackers will not compromise the conference website and thus, the bank may not need to alter traffic to it. Identifying such websites in pre-processing could greatly reduce the size of our problem. 
A website $w$ is \emph{dominated by another website} $u$ if the attacker would not attack $w$ unless they have used the maximum effort on $u$, i.e. $e_u = t_u^{all}$, regardless of the defender's strategy. 
Thm \ref{thm:elimination} presents sufficient conditions for a website to be dominated and leads to an algorithm (Alg. \ref{algline:elimcondition}) to find dominated website to be eliminated.
\begin{theorem} \label{thm:elimination}
Consider websites $u, w \in W$. If the following conditions hold, the website $w$ is dominated by $u$:
\begin{align*}
& x_u^{max} := B_d/(c_u t_u) \leq 1, & \kappa_w & \leq \kappa_u (1- x_u^{max}),\\
& \pi_w \geq \pi_u, & t_w^{all} & \leq t_u^{all}.
\end{align*}
\end{theorem}
\begin{algorithm}[t] 
	Define $U = \{w \in W: c_w t_w \geq B_d\}$. Let $D = \emptyset$.\\
	Calculate $x_u^{max} = B_d/c_u t_u,\forall u\in U$\\
	\ForEach{website $w \in W$}{
    	Set $U_w = \{u \in U: \kappa_w \leq \kappa_u(1- x_u^{max})\}$\\
    	\lIf{exists $U_w^* \subseteq U_w$ such that \\
        (1) $\sum_{u \in U_w^*} \pi_u \leq \pi_w$, 
        (2) $\sum_{u \in U_w^*} t_u^{all} \geq t_w^{all}$, 
        and (3) $\sum_{u \in U_w^*} t_u^{all} \geq B_e$}{ \label{algline:elimcondition}
        $D = D \cup \{w\}$
        }
    }
    \Return set of dominated websites $D$ 
	\caption{\textsc{Find-dominated-websites}}
	\label{alg:elim}
\end{algorithm}

We conclude the section with the following claim.
\begin{claim} \label{clm:iaopt}
\textsc{CyberTWEAK} terminates with optimal solution.
\end{claim}
In light of the hardness of the attacker's best response problem (Thm \ref{thm:abr}), we also design a variant of \textsc{CyberTWEAK}, which uses a greedy heuristic to find a new max effort vector to be added in each iteration of column generation (denoted as \textsc{GreedyTWEAK}). The algorithm allocates the adversary's budget to websites in decreasing order of $r_w = \kappa_w(1-x_w)\alpha_w$, where $\alpha_w$ is a tuning parameter. Another variant uses an exact dynamic programming algorithm for the slave problem. Details about these variants can be found in Appendix~\ref{app:algs}. Also, we note that the SED problem is related to the recent work on bi-level knapsack with interdiction~\cite{caprara2016bilevel}. However, our outer problem of $\mathcal P_1$ is continuous rather than discrete, and the added dimension of adversary's effort makes the inner problem $\mathcal P_2(x)$ more complicated than that being studied in this work.

\section{Experiments}
We developed and tested \textsc{CyberTWEAK} to match the scalability required of large-scale deployment. 
Unless otherwise noted, problem parameters are described in details in Appendix~\ref{app:params}. 
All results are averaged over 20 instances; error bars represent standard deviations of the mean. 

\begin{figure}[t]
    \subfloat[Tractable cases]{\includegraphics[width=0.5\columnwidth]{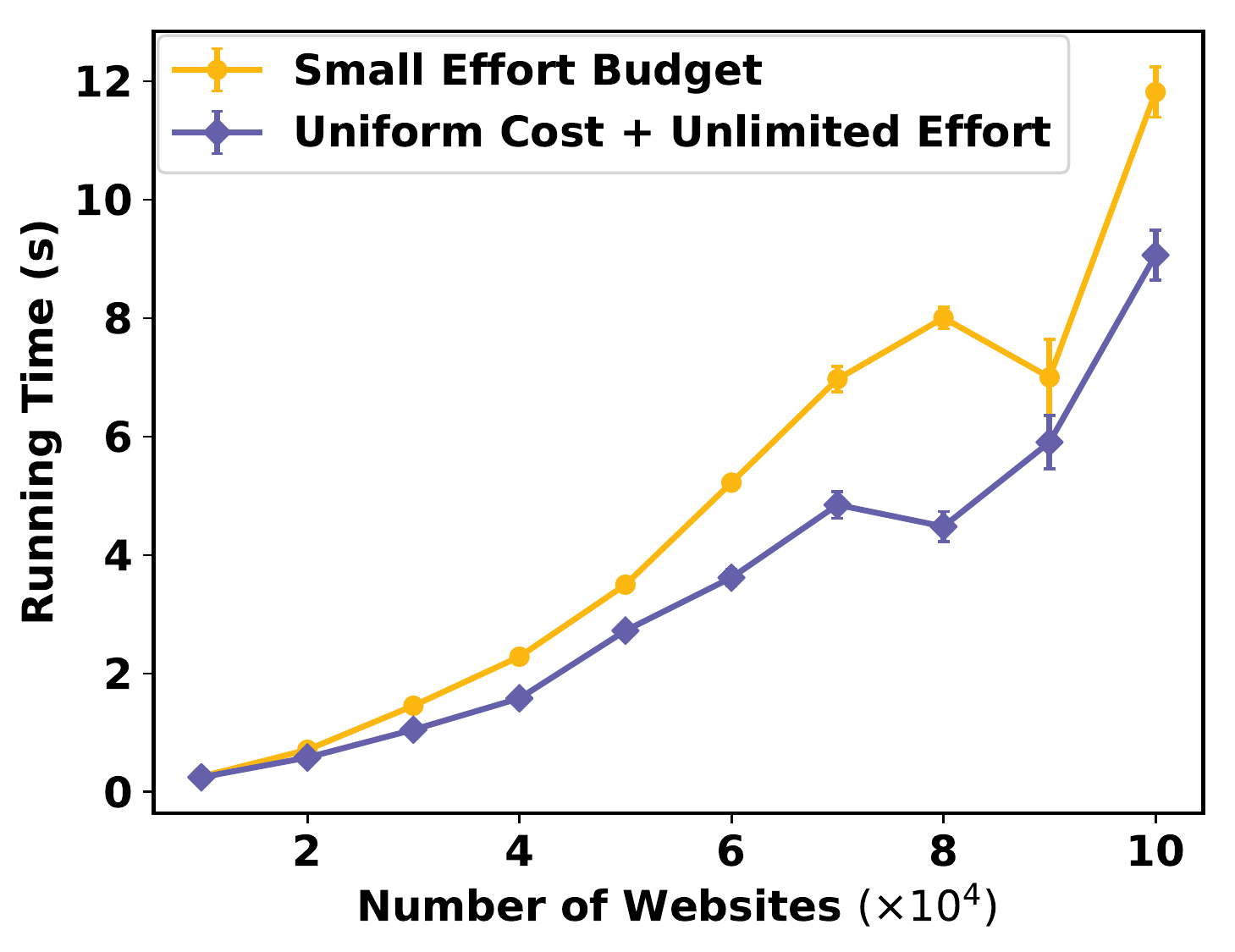}%
        \label{fig:easy}}
     \subfloat[Small instances]{\includegraphics[width=0.5\columnwidth]{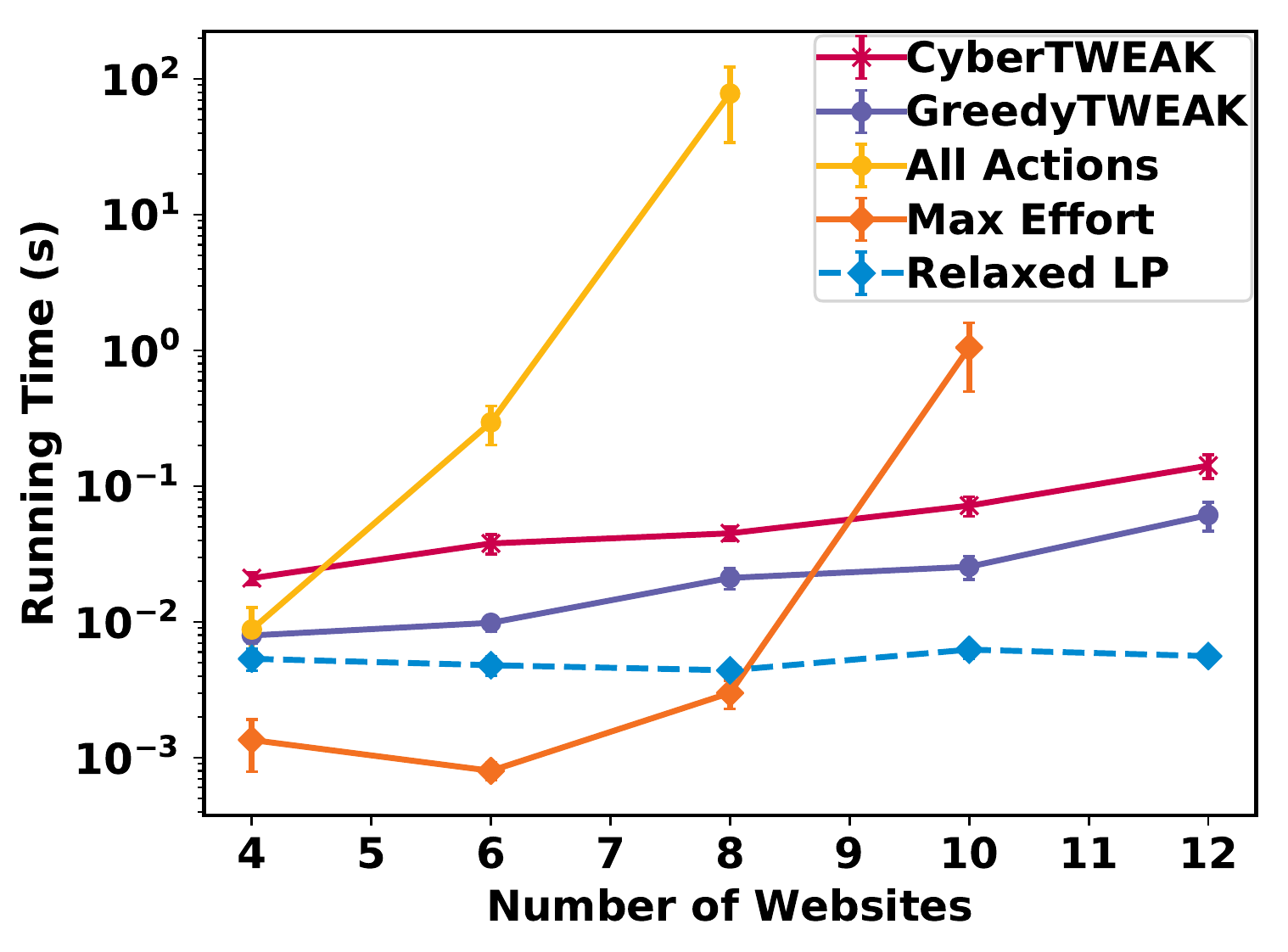}%
        \label{fig:small}}\\
   \subfloat[Medium instances running time]{\includegraphics[width=0.5\columnwidth]{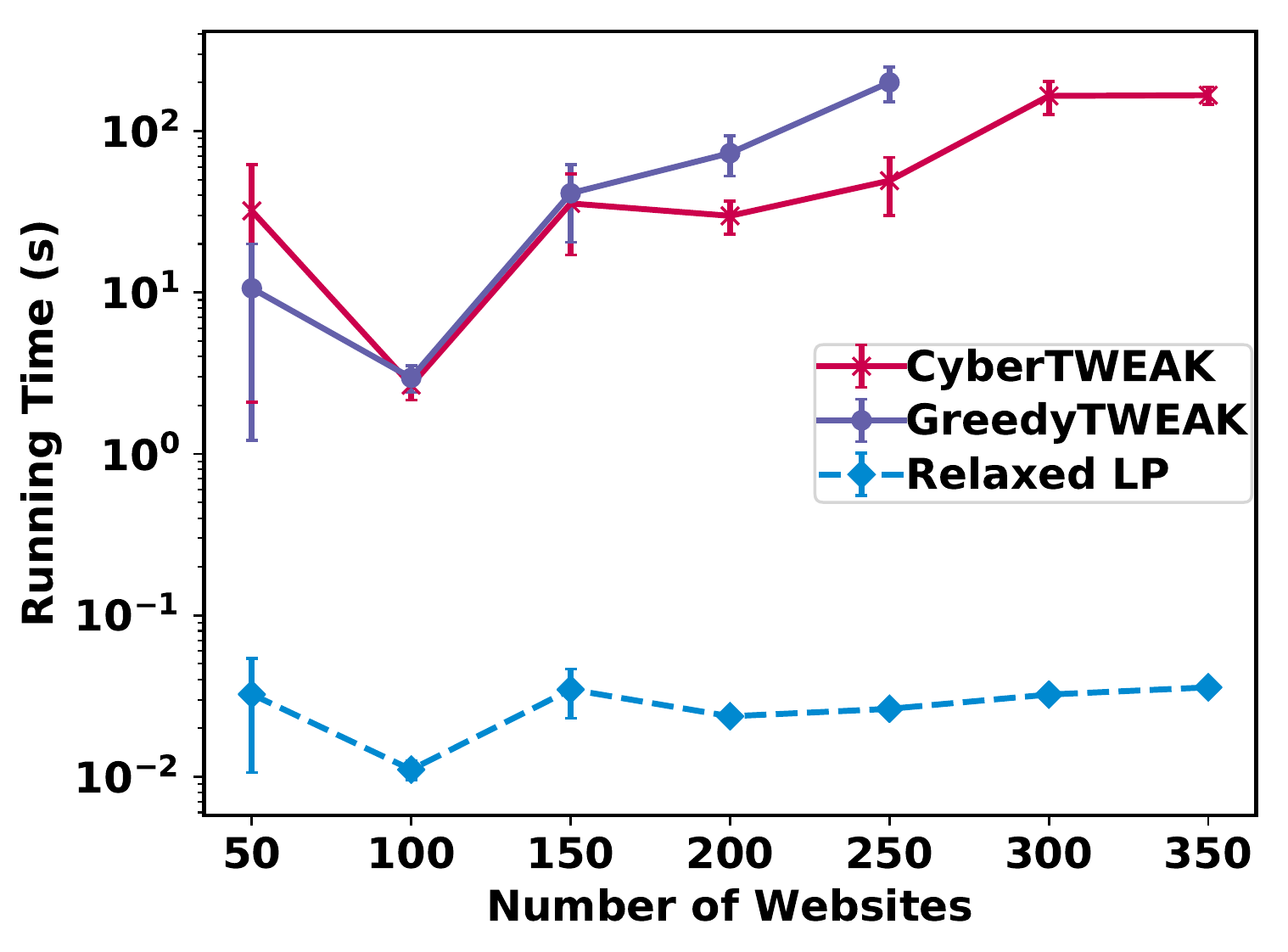}
        \label{fig:large}}
     \subfloat[Medium instances \#strategies]{\includegraphics[width=0.5\columnwidth]{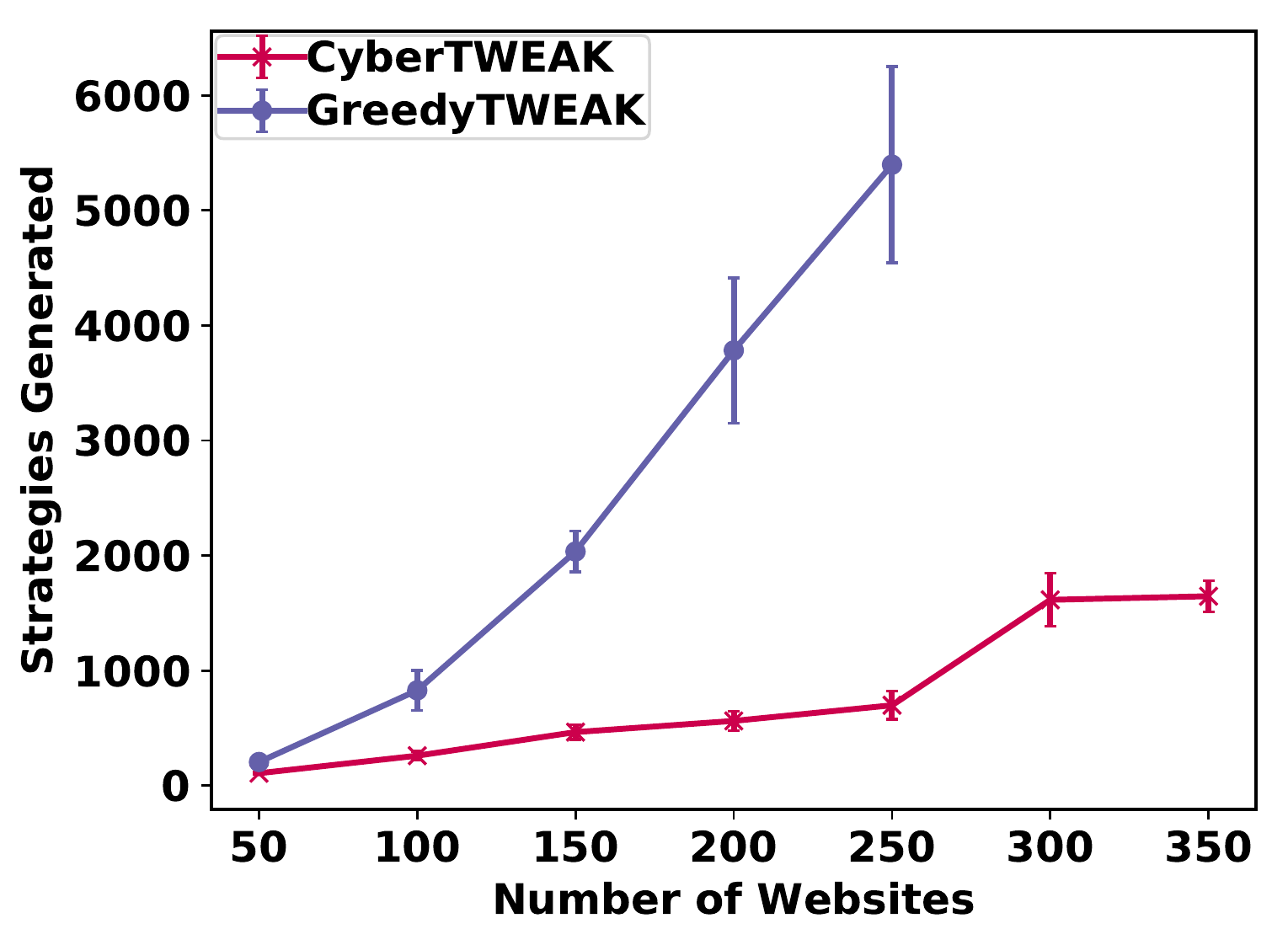}
        \label{fig:largeiter}}\\
    \subfloat[Large instances]{\includegraphics[width=0.5\columnwidth]{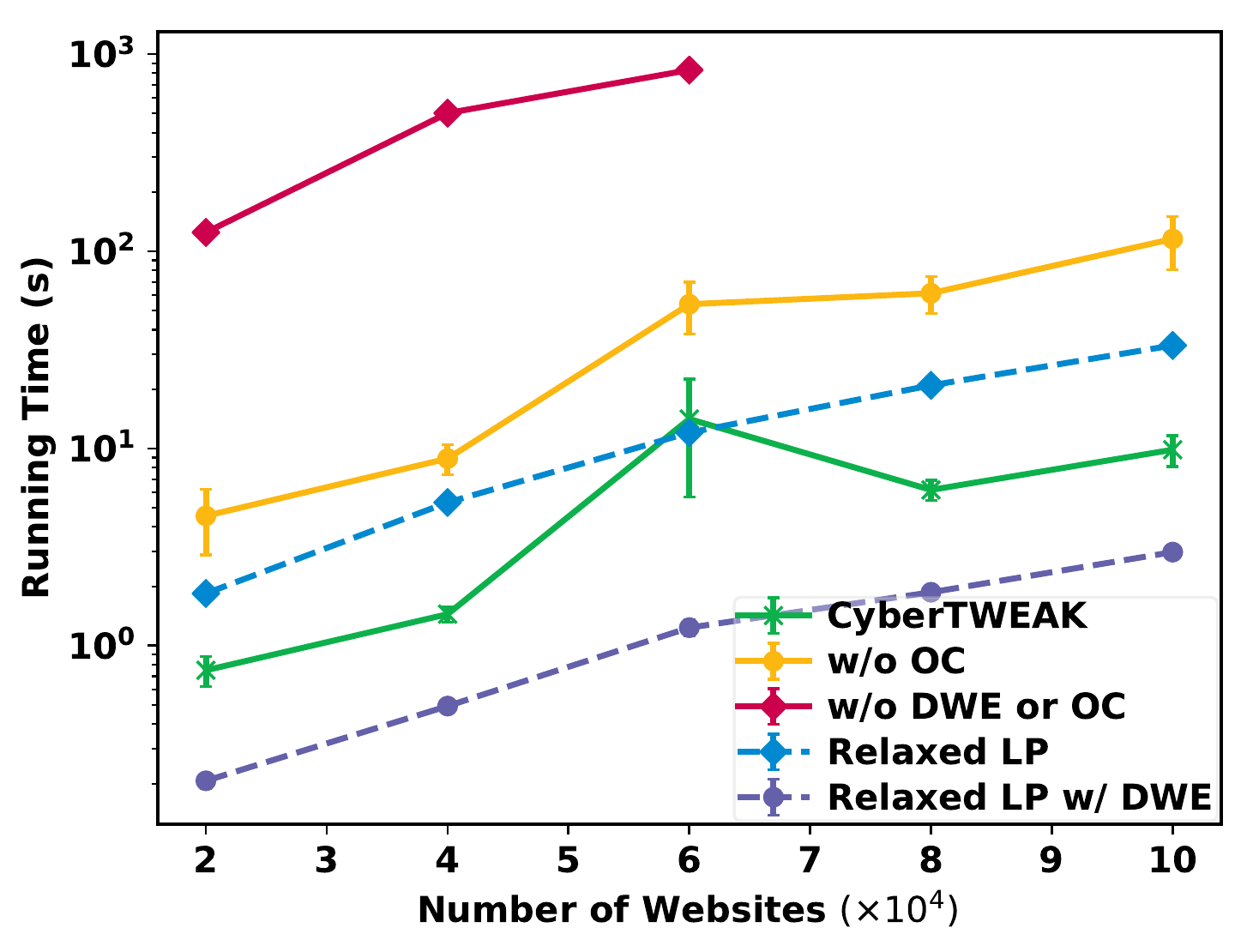}\,
        \label{fig:preprocess}}
    \subfloat[Trade-off]{\includegraphics[width=0.5\columnwidth]{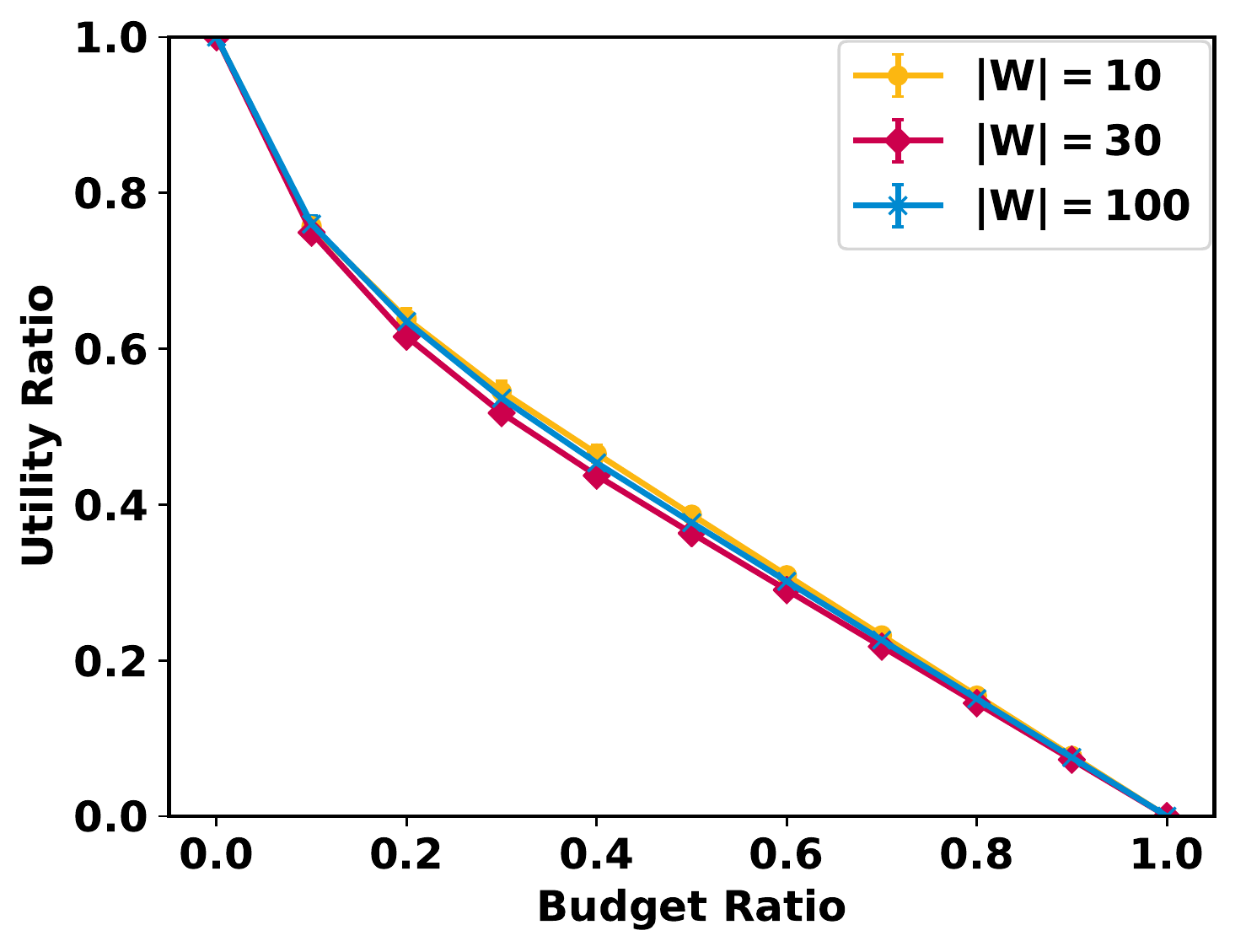}
        \label{fig:tradeoff}} 
    \caption{Experiment results}
    \label{fig:all}
\end{figure}

First, we run experiments on the polynomial time tractable cases (Corollary~\ref{cor:smallbudget} and Theorem~\ref{thm:unifcost}). Fig.~\ref{fig:easy} shows that in both cases, our solution can easily handle $10^5$ websites, applicable to real-world corporate-scale problems.

Moving on to the general SED games, we test 3 algorithms (\textsc{CyberTWEAK}, \textsc{GreedyTWEAK}, and \textsc{RelaxedLP}) with two other baselines, \textsc{MaxEffort} and \textsc{AllActions}. 
\textsc{RelaxedLP} refers to solving $\hat{\mathcal P_1}$.
\textsc{MaxEffort} solves $\mathcal P_1^{\text{LP}}(\hat e^{\mathcal A})$ directly without column generation. \textsc{AllActions} decomposes SED into subproblems, each assuming some adversary's effort vector is a best response. Its details can be found in Appendix~\ref{app:algs}. 
We test the algorithms with different problem scales. In small and medium sized instances, we skip dominated website eliminateion (DWE) step (Line \ref{algline:elimcondition}) and optimality check (OC) step (Line \ref{algline:checkopt}) in Alg. \ref{alg:cutgen} as the problem size is small enough, making these steps unnecessary. We use solid lines to represent methods with optimality guarantee and dotted lines for others (\textsc{RelaxedLP} based methods).

For small instances (Fig.~\ref{fig:small}), both baselines become impractical even on problems with less than $12$ websites. However, \textsc{CyberTWEAK} is able to find the optimal solutions rather efficiently. \textsc{GreedyTWEAK} slightly improves over \textsc{CyberTWEAK}. \textsc{RelaxedLP} yields the fastest running time, despite a solution gap above $6\%$ as shown in Table~\ref{tab:relaxedlperror}.

\begin{table}[t]
\centering
    \begin{tabular}{cccccc}\hline
        $|W|$  & Gap & \# Exact &  $|W|$  & Gap & \# Exact \\\hline
        4 & $13.19\%$ & 2/20 & 150 & 7e-8 & 16/20\\ 
        8 & $8.11\%$ & 5/20 & 200 & 8e-10 & 19/20\\
        12 & $6.63\%$ & 8/20 & 250 & $0$ & 20/20\\
        50 & 2e-6 & 18/20 & 300 & 2e-3 & 17/20\\
        100 & 8e-9 & 19/20 & 350 & 2e-8 & 18/20\\\hline
    \end{tabular}
    \caption{Solution quality of \textsc{RelaxedLP}, with the number of instances where \textsc{RelaxedLP} solves the problem exactly.}
    \label{tab:relaxedlperror}
\end{table}

For medium-sized instances (Fig.~\ref{fig:large}), baseline algorithms cannot run and \textsc{GreedyTWEAK} stops being helpful,
mainly because the ``better'' effort vectors generated in \textsc{GreedyTWEAK} far outnumbers the ``best'' effort vectors in \textsc{CyberTWEAK} (Fig.~\ref{fig:largeiter}) despite the saved time in each iteration.
\textsc{Relaxed LP} has negligible running time and often solves the problem optimally (Table~\ref{tab:relaxedlperror}). 

For large instances (Fig.~\ref{fig:preprocess}), \textsc{CyberTWEAK} with both DWE and OC steps 
is able to handles $10^5$ websites in 10 seconds. 
When we remove (denoted as ``w\textbackslash o'') DWE and/or OC step, runtime increases significantly, showing the efficacy of these steps\footnote{The impact of DWE varies significantly across instances and relies heavily on the distribution of traffic. In less than 4 of the 20 instances DWE did not reduce the problem size by much. We report in Fig.~\ref{fig:preprocess} the majority group where DWE eliminated a significant number of websites. We provide further discussion in Appendix~\ref{app:params}.}
Compared to \textsc{RelaxedLP} or \textsc{RelaxedLP} enhanced with DWE step, which can also efficiently handles $10^5$ websites, \textsc{CyberTWEAK} has optimality guarantee. 

Finally, we consider the trade-off between the risk exposure and degradation in rendering websites, represented by the objective $OPT(\mathcal P_1)$ and defender's budget $B_d$, respectively. With budget $\bar B_d = \sum_{w \in W} c_w t_w$, the attacker would have zero utility. With zero defender budget, the attacker would get maximum utility $\bar U$. Fig.~\ref{fig:tradeoff} shows how the utility ratio $OPT(\mathcal P_1)/\bar U$ changes with the budget ratio $B_d/\bar B_d$. As the organization increases the tolerance for service degradation, its risk exposure drops at a decreasing rate.

\section{Deployment}
Based on \textsc{CyberTWEAK}, we developed a browser extension (available on the Google Chrome Web Store\cref{fnt:extension}). It can modify the user-agent string sent to websites automatically during browsing which contains information such as the operating system, browser, and services running on the user's machine. 
The extension receives from the user the websites visited $W$, number of visits per week $t_w$, the cost to alter the user-agent string $c_w$ and budget $B_d$. The total traffic $t_w^{all}$ and attack cost $\pi_w$ are estimated from the Cisco Umbrella 1 Million list~\cite{cisco2019}. The attacker's budgets are set in scale with the previously mentioned parameters. The extension runs \textsc{CyberTWEAK} to set the probability of altering the user-agent string for each website. Note that it is the relative magnitudes, rather than the exact values, that matter. 

\begin{figure}[t]
    \centering
    \includegraphics[width=0.9\columnwidth, height=1.5in]{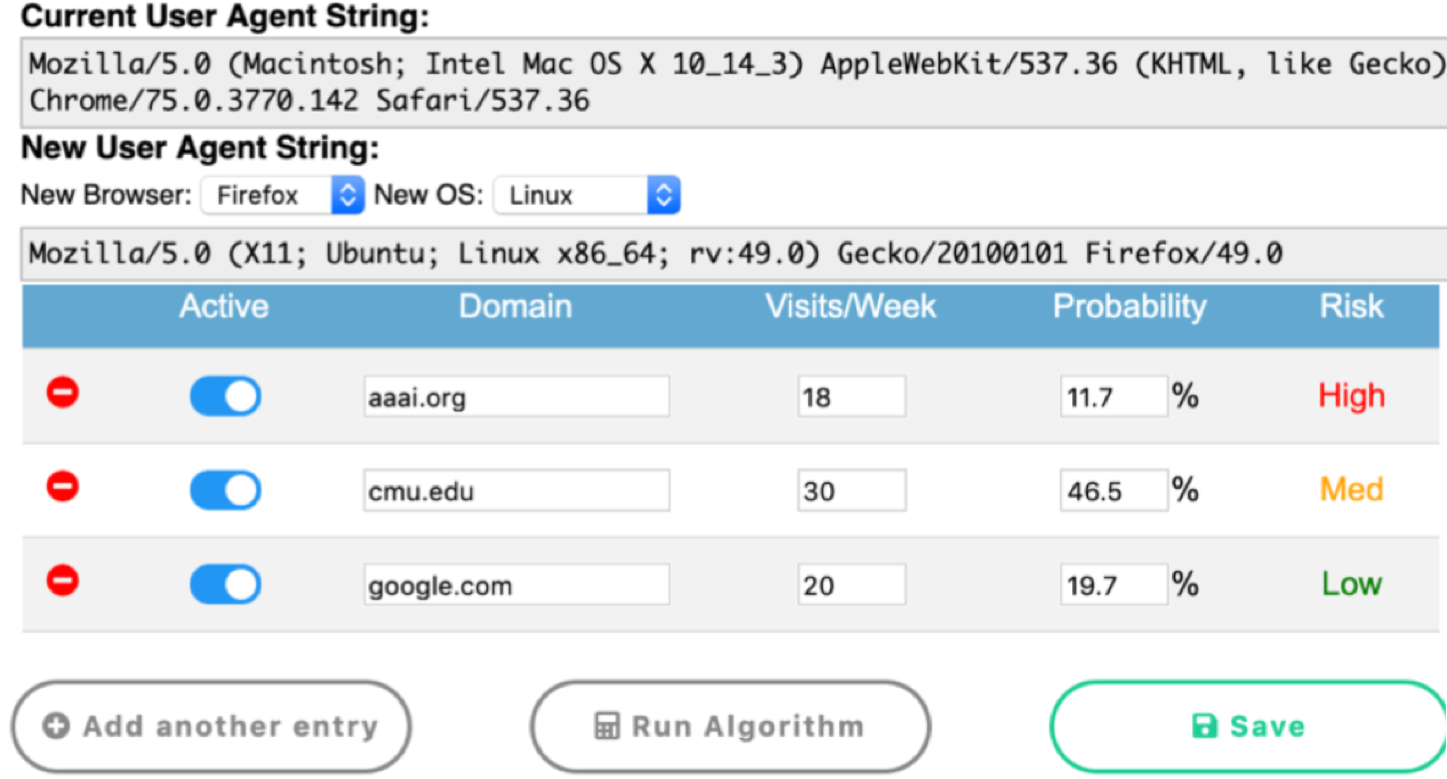} 
    \includegraphics[width=0.8\columnwidth, height=1.5in]{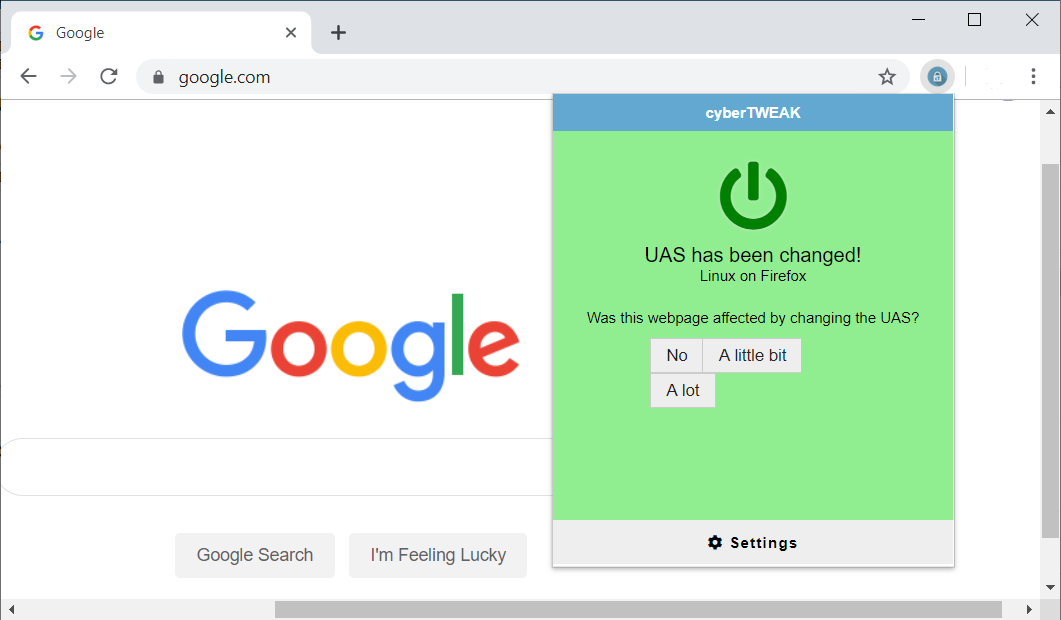}
    \caption{Screenshots of the browser extension}
    \label{fig:extension}
\end{figure}

The extension takes additional steps to make our algorithm more usable and interpretable. First, some users may find it hard to specify the cost of altering user-agent string $c_w$ and budget $B_d$. Our extension will adjust the values based on the qualitative feedback provided by users about whether the degradation of the website's rendering is acceptable when they visit a website using the modified user-agent, as shown in Fig.~\ref{fig:extension}.
Second, in addition to showing the computed altering probabilities, the extension also displays a personalized ``risk level'' for each website, to help the user understand the algorithm's output. Less popular websites  frequented more often by the user have higher risk, as shown in Fig.~\ref{fig:extension}.

As mentioned in Section~\ref{sec:model}, advanced cyber attackers might sometimes circumvent the existing deception methods. Future versions of the extension will leverage the latest advances in anti-fingerprinting techniques, which entail manipulating more than the user-agent string. 

We believe this \textsc{CyberTWEAK} extension is vital to the continued study and development of the countermeasure we develop for this domain and large scale deployments.

\section*{Acknowledgments}
Co-authors Z. R. Shi and F. Fang are supported in part by the U.S. Army Combat Capabilities Development Command Army Research Laboratory under Cooperative Agreement Number W911NF-13-2-0045 (ARL Cyber Security CRA). 

{\fontsize{9.0pt}{10.0pt}\selectfont
\bibliographystyle{aaai}
\bibliography{aamas19}
}
\appendix
\begin{center}
\LARGE\textbf{Draining the Water Hole: \\Mitigating Social Engineering Attacks with CyberTWEAK}\\
\bigskip
\LARGE\textbf{Appendix}
\end{center}

\bigskip
\section{Deferred Algorithms}\label{app:algs}
\subsection{Attacker's Better Response Heuristic}
In light of the hardness of finding the adversary's best response, we consider a greedy heuristic. Leveraging Theorem~\ref{thm:greedy-allocation}, \textsc{Greedy} (Alg.~\ref{alg:greedy}) allocates the adversary's budget to websites in decreasing order of the ratio $r_w = \frac{t_w(1-x_w)/t^{all}_w}{\alpha_w}$, where $\alpha_w$ is a tuning parameter.
We replace the MILP for $\mathcal P_2(x)$ in \textsc{CyberTWEAK} with Alg.~\ref{alg:greedy} to find an adversary's better response. If it does not yield a new effort vector, the MILP is called. The column generation process terminates if the MILP again does not find a new effort vector. We refer to this entire procedure as \textsc{GreedyTWEAK}. Note that \textsc{GreedyTWEAK} also terminates with the optimal solution.
Although \textsc{Greedy} (Alg.~\ref{alg:greedy}) does not provide an approximation guarantee, it performs well in practice. 
As we show in the experiment section, in practice the accuracy of its solution improves as the size of the problem grows.
We also considered a dynamic programming algorithm which is exact and runs in pseudo-polynomial time. However, its practical performance is unsatisfactory.
\begin{algorithm}
	Sort the websites in decreasing order of $r_w = \frac{t_w(1-x_w)/t^{all}_w}{\alpha_w}$.\\
	\ForEach{website $w$ in the sorted order}{
		\If{remaining attack budget $\geq$ attack cost $\pi_w$}{
			Attack this website $w$ with maximum effort allowed
		}
		\lIf{running out of budget}{
			\textbf{break}
		}
	}
	\caption{\textsc{Greedy}}\label{alg:greedy}
\end{algorithm}

\subsection{Baseline Algorithm for $\mathcal P_1$}
\begin{algorithm}[t]   
	\ForEach{$(a^{*}, w^*) \in \mathcal{A} \times W$ where $w^* \in a^*$}{
    	\ForEach{website $w \in W$}{
    	    \uIf{$w=w^*$}{
    	        Define $z_w=\min\{B_e-\sum_{w\in a^{*}, w\neq w^*}t_{w}^{all},t_{w^*}^{all}\}$
    	    }
    	    \uElseIf{$w \in a^*$}{
    	        Define $z_w = t_{w}^{all}$
    	    }
    	    \Else{
    	        Define $z_w = 0$
    	    }
        	 Define $k_w=\frac{t_w}{t_w^{all}}z_w$\\
        	 \ForEach{$(\hat a, \hat w) \in \mathcal{A} \times W$ where $\hat w \in \hat a$}{
        	 Define $\hat k_w$ similarly as above, for each $w \in W$.\\
        	 Add to $BR(a^*, w^*)$ the following linear constraint
$\sum_{w\in a^{*}} k_w(1-x_w) \geq \sum_{w\in \hat{a}} \hat{k}_w(1-x_w)$\\
        }
        }
        	 Solve the following LP
\begin{align}
\min_{x, v} & \quad v \label{lp9Eq1}\\
\ \ \  \text{s.t.} & \quad  v \geq  \sum_{w\in W} k_w(1-x_w)  \label{lp9Eq2} \\
&\quad \mbox{linear constraints in $BR(a^*, w^*)$} \label{lp9Eq3} \\
& \quad \sum_{w\in W} c_w t_w x_w \leq B_d \label{lp9Eq4} \\ 
&\quad x_w\in [0,1],\qquad \forall w\in W  \label{lp9Eq5}
\end{align}
        }
        Select the best solution out of all the LPs.
	\caption{\textsc{All Actions}}
	\label{alg:allactions}
\end{algorithm}

We show the details of one of our baseline algorithms, All Actions, in Alg.~\ref{alg:allactions}. Let $\mathcal{A}$ denote the set of actions available to the adversary such that the budget constraint is satisfied. Each action $a^* \in \mathcal A$ is a set of websites being compromised. According to Theorem~\ref{thm:greedy-allocation}, among all the websites $w$ compromised in $a^*$, the adversary puts ``partial'' effort $e_w \in (0, t_w^{all}$ on at most one website $w^*$. Therefore, the action-website pairs $(a^*, w^*)$ fully characterize the adversary's strategies. Alg.~\ref{alg:allactions} works by finding the optimal defender strategy, assuming each action-website pair is the optimal strategy for the adversary.

\section{Deferred Proofs}
\subsection{Proof of Theorem~\ref{thm:abr}}
\label{app:abr}
We reduce from the knapsack problem. 
In the knapsack problem, we have a set $W$ of items each with a weight $\omega_w$ and value $p_w$ $\forall w\in N$, and aim to pick items of maximum possible value subject to a capacity $B$. 
We now create an instance of the SED problem. 
Create a website for each item $w \in W$ with organization traffic and total traffic $t_w = t^{all}_w = p_w$ and attack cost $\omega_w$. 
Assume that $x = {\bf 0}^T$. 
Next, set $B_a = B$ and $B_e = \infty$. 
Notice that the objective function becomes $\sum_{w \in W} e_w$ where $\sum_{w \in W} e_w \leq \infty$ and $e_w \leq p_w y_w$. 
Hence, $e_w = p_w$ whenever $y_w=1$. 
Then, the adversary's best response problem is given by:
\begin{align}
\max_{y} & \quad  \sum_{w \in W} p_w y_w & \label{lp3Eq1}\\
\ \ \  \text{s.t.} & \quad \sum_{w \in W} \omega_w y_w \leq B &   \label{lp3Eq2} \\
&\quad y_w\in \{0,1\} & \forall w\in W  \label{lp3Eq3}
\end{align}
This is exactly the knapsack problem described above.
\qed

\subsection{Proof of Theorem~\ref{thm:greedy-allocation}}
\label{app:greedy-allocation}
	For each $w \in W$, let $k_w = t_w (1-x^*_w)/t_w^{all}$. Suppose there exist some $w_1, w_2 \in W_B$, and w.l.o.g assume $k_{w_1} \geq k_{w_2}$. Let $\Delta e = \min\{e^*_{w_2}, t_{w_1}^{all} - e_{w_1}^*\}$. Consider the solution $(x^*, y^*, \hat e)$ where $\hat e_{w_1} = e_{w_1}^* + \Delta e$, $\hat e_{w_2} = e_{w_2}^* - \Delta e$, and $\hat e_w = e^*_w$ for all other websites $w \in W$. This is a feasible solution, and the objective increases by $(k_{w_1} - k_{w_2}) \Delta e \geq 0$ compared to $(x^*, y^*, e^*)$. Furthermore, at least one of $w_1$ and $w_2$ is removed from $W_B$. We can apply this argument repeatedly until $|W_B| \leq 1$.
	\qed

\subsection{Proof of Corollary~\ref{cor:smallbudget}}
\label{app:smallbudget}
	Since $B_e \leq t^{all}_w \ \ \forall w\in W$, we know $|W_F| \leq 1$ for any feasible solution. If $|W_F| = 1$, then we have $|W_Z| = n-1$ and $|W_B| = 0$. If $|W_F| = 0$, by Theorem~\ref{thm:greedy-allocation}, we have $|W_B| = 1$ and $|W_Z| = n-1$. In either case, there is only website $w^*$ such that $e_{w^*} > 0$. It follows that $w^* \in \arg\max_{w\in W} \frac{t_w(1-x_w)B_e}{t^{all}_w}$ given a defender strategy $x$. The optimal defender strategy can be found by solving the following LP.
\begin{align}
\min_{x, v} & \quad v & \label{lp4Eq1}\\
\ \ \  \text{s.t.} & \quad  v \geq  \frac{t_w(1-x_w)B_e}{t^{all}_w} & \forall w\in W  \label{lp4Eq2} \\
&\quad \sum_{w\in W} c_w t_w x_w \leq B_d &   \label{lp4Eq3} \\
&\quad x_w\in [0,1] & \forall w\in W  \label{lp4Eq4} \qed
\end{align}

\subsection{Proof of Theorem~\ref{thm:unifcost}}
\label{app:unifcost}
Under these assumptions, the problem $\mathcal P_1$ becomes
\begin{align}
\min_{x} \max_{y,e} & \quad  \sum_{w\in W} t_w(1-x_w) y_w &\\
\ \ \  \text{s.t.} &\quad \sum_{w\in W} y_w \leq B_a & \\
& \quad   \sum_{w\in W} c_w t_w x_w \leq B_d &  \\
&\quad x_w\in [0,1], y_w\in \{0,1\} & \forall w\in W 
\end{align}

The constraint $\sum_{w\in W} y_w \leq B_a$ must be satisfied with equality because $t_w (1-x_w) \geq 0$ for all $w \in W$. The defender's problem is to minimize the sum of $B_a$ largest linear functions $t_w - t_w x_w$ among the $n = |W|$ of them, subject to the polyhedral constraints on $x_w$. This problem can be solved as a single LP
(Ogryczak and Tamir 2003) as follows.
\begin{align}
\min_{d^+, x, z}  & \quad  B_a z + \sum_{w \in W} d_w^+ &\\
\ \ \  \text{s.t.} &\quad d_w^+ \geq t_w - t_w x_w - z & \forall w\in W \\
& \quad   \sum_{w \in W} c_w t_w x_w \leq B_d &  \\
&\quad x_w\in [0,1],\, d_w^+ \geq 0 & \forall w\in W 
\qed
\end{align}

\subsection{Proof of Theorem~\ref{thm:3approx}}
\label{app:3approx}
Let $x^*$ be the optimal solution to $\mathcal P_1$. Consider the problem $\hat{\mathcal P_2}(x^*)$. 
At optimal solution, the inequality $e_w\leq t^{all}_w \cdot y_w$ in $\hat{\mathcal P_2}(x^*)$ is satisfied with equality, as if $e_w< t^{all}_w \cdot y_w$, then we can decrease $y_w$ without changing the objective value and violating any constraints. Then, we can eliminate the variables $e_w$ and $\hat{\mathcal P_2}(x^*)$ becomes a standard two-dimensional fractional knapsack problem $\hat{\mathcal P_4}(x^*)$. It is well-known that there exists an optimal solution to $\hat{\mathcal P_4}(x^*)$ which has at most 2 fractional values $y_{w_1}$ and $y_{w_2}$ (Kellerer, Pferschy, and Pisinger 2004).
We have
\[
\begin{split}
& OPT(\hat{\mathcal P_1}) \leq OPT(\hat{\mathcal P_2}(x^*)) = OPT(\hat{\mathcal P_4}(x^*))\\
& \leq OPT(\mathcal P_2(x^*)) + t_{w_1} (1-x^*_{w_1}) + t_{w_2} (1-x^*_{w_2}) \\
& \leq 3 OPT(\mathcal P_2(x^*)) = 3 OPT(\mathcal P_1)
\end{split}
\]
Note that if $B_e = \infty$, $\hat{\mathcal P_1}$ is a 2-approximation.
\qed

\subsection{Proof of Theorem~\ref{thm:opt}}
\label{app:opt}
Since $\hat{x}^*$ and its best response calculated by $\mathcal P_2(\hat{x}^*)$ form a feasible solution to $\mathcal P_1$, the first inequality holds. 
For any defender strategy $x$, $\text{OPT}(\mathcal P_2(x)) \leq \text{OPT}(\hat{\mathcal P_2}(x))$ as adversary can choose fractional $y_w$'s in $\hat{\mathcal P_2}(x)$. For $\hat{x}^*$ specifically, we have $\text{OPT}(\hat{\mathcal P_2}(\hat{x}^*)) = \text{OPT}(\hat{\mathcal P_1})$, since $\hat{\mathcal P_1}$ is, by strong duality, equivalent to $\mathcal P_1$ except that the adversary is allowed to choose fractional $y_w$'s. This establishes the second inequality.
The last inequality holds because $x^*$ and its fractional best response calculated by $\hat{\mathcal P_2}(x^*)$ form a feasible solution to $\hat{\mathcal P_1}$.
\qed

\subsection{Proof of Theorem~\ref{thm:elimination}}
\label{app:elimination}
From conditions (1) and (2), we know that for the same amount of effort, the attacker will be better off attacking website $u$ than $w$, regardless of the defender's strategy. 

Suppose $e_w > 0$ and $e_u = 0$ (consequently $y_w = 1, y_u = 0$). Then we could let $e_w' = 0$ and $e_u' = e_w$. This is possible because from condition (4), $e_w \leq t_{w}^{all} \leq t_u^{all}$ so we have $e_u' \leq t_u^{all}$. Doing this does not increase the attack cost because now $y_w' = 0$ and $y_u' = 1$ and $\pi_w \geq \pi_u$ from condition (3). 

Suppose $e_w > 0$ and $e_u > 0$ (consequently $y_w = y_u = 1$). Let $e_w' = e_w - \min\{e_w, t_u^{all} - e_u\}$ and $e_u' = e_u + \min\{e_w, t_u^{all} - e_u\}$. We know that if $e_w' > 0$, then $e_u' = t_u^{all}$. Of course, the attack cost does not increase as well.
\qed

\subsection{Proof of Claim~\ref{clm:lpopt}}
\label{app:lpopt}
Suppose $(\hat{x}^*, OPT(P_2(\hat{x}^*)))$ is not an optimal solution for the LP $\mathcal P_1^{\text{LP}}(\hat e^{\mathcal A})$ which is equivalent to $\mathcal P_1$. Thus, equivalently $\hat{x}^*$ not optimal for $\mathcal P_1$. Any of its neighborhood with radius $\epsilon$ contains some $(\hat{x}', v')$ as a better solution, meaning $v' < OPT(P_2(\hat{x}^*))$. This solution $(\hat{x}', v')$ satisfies constraint~\eqref{lp7Eq2}, which is strictly stronger than constraint~\eqref{lp8Eq2}. Therefore $(\hat{x}', v')$ is feasible for $\tilde P_3(\hat{x}^*)$; this contradicts $OPT(\tilde P_3(\hat{x}^*)) \geq OPT(P_2(\hat{x}^*))$. 
\qed

\subsection{Proof of Claim~\ref{clm:iaopt}}
\label{app:iaopt}
Claim~\ref{clm:lpopt} has covered the case where \textsc{CyberTWEAK} terminates after the optimality check on Line \ref{algline:checkopt}, Alg.~\ref{alg:cutgen}. In the other case, \textsc{CyberTWEAK} terminates when no new effort vectors are found for the adversary. Suppose $x$ is the optimal solution to the defender's optimization problem (Line~\ref{algline:solvelp}, Alg.~\ref{alg:cutgen}), and suppose now $\mathcal P_2(x)$ does not find a new effort vector (Line~\ref{algline:attall}, Alg.~\ref{alg:cutgen}). This implies $x$ would still be feasible for the LP $\mathcal P_1^{\text{LP}}( e^{\mathcal A})$ even if $e^{\mathcal A}$ is replaced by the set of all max effort vectors $\hat e^{\mathcal A}$. Thus, $x$ is an optimal solution. Indeed, at this point the optimal values of $\mathcal P_1^{\text{LP}}( e^{\mathcal A})$ and $\mathcal P_2(x)$ are equal.
\qed

\section{Deferred Experiments}\label{app:experiments}
We present additional experiments on the adversary's best response problem.
In the \textsc{greedy} algorithm (Alg.~\ref{alg:greedy}), the adversary selects websites based on a decreasing order of $r_w = \frac{t_w(1-x_w)/t_w^{all}}{\alpha_w}$. Here, $\alpha_w$ is the tuning parameter.  
With different choices of $\alpha_w$, we compare the output value $\text{OPT}_{\text{Greedy}}$ of \textsc{Greedy} with the optimal value $\text{OPT}$ obtained by solving the MILP $\mathcal P_2(x)$. Table~\ref{tab:greedychoice} shows the solution gap $\frac{\text{OPT} - \text{OPT}_{\text{Greedy}}}{\text{OPT}}$. We observe that $\alpha_w = \pi_w$ yields the smallest solution gap. We also tested other choices for $\alpha_w$ such as $(\pi_w/B_a)^p + (1/B_e)^q$ for different powers $p$ and $q$, yet they do not yield better optimization gaps. Hence we fix $r_w = \frac{t_w(1-x_w)/t_w^{all}}{\pi_w}$ in subsequent experiments. 
\begin{table}[]
    \centering
    \begin{tabular}{cc}\hline
        $\alpha_w$  & $\frac{\text{OPT} - \text{OPT}_{\text{Greedy}}}{\text{OPT}}$ \\\hline
        $\pi_w$ & $0.0079$\\ 
        $\pi_w/B_a + 1/B_e$ & $0.0285$\\
        $1$ & $0.0082$\\\hline
    \end{tabular}
    \caption{Solution gaps of different greedy heuristics for the adversary best response problem. Results are averaged over 5 runs on different problem sizes $|W| = 100, 200, \dots, 500$.}
    \label{tab:greedychoice}
\end{table}

\begin{figure}[t]
     \subfloat[\textsc{Greedy} running time]{\includegraphics[height=1.1in, width=0.5\columnwidth]{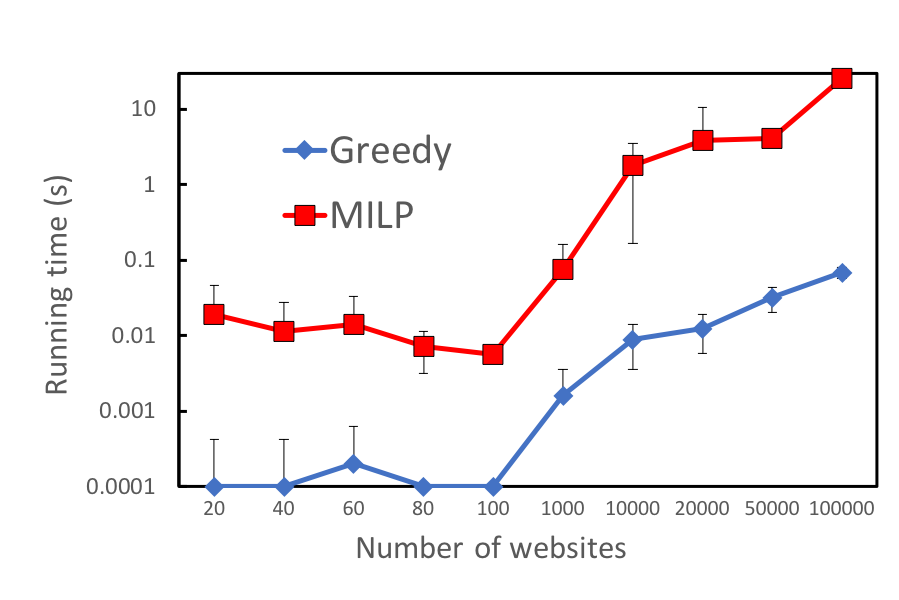}%
        \label{fig:greedy}} 
    \subfloat[\textsc{Greedy} solution gap]{\includegraphics[clip, trim=0.00in 0.3in 0.00in 0.00in, height=1.05in, width=0.5\columnwidth]{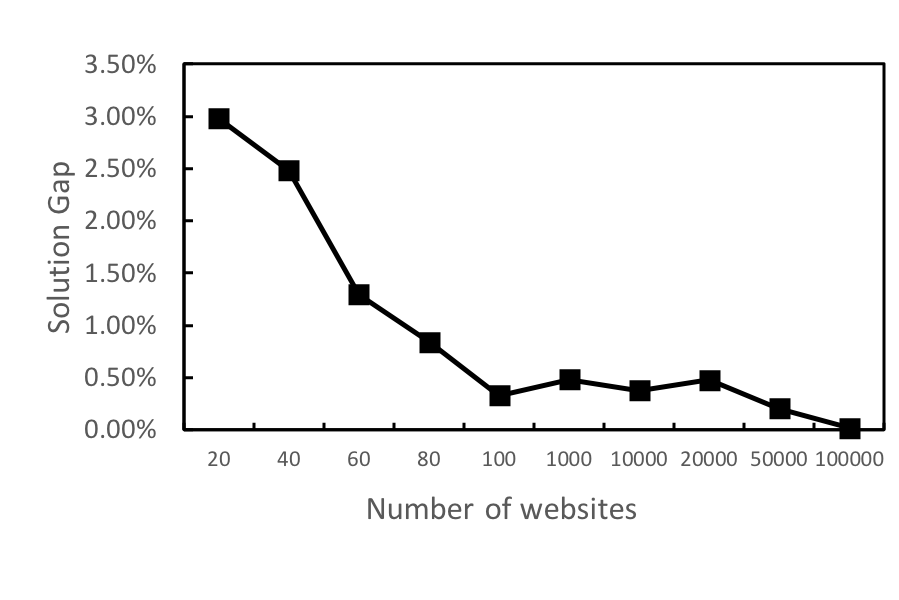}%
        \label{fig:greedygap}}
\end{figure}

Fig.~\ref{fig:greedygap} shows \textsc{Greedy}'s solution gap decreases to near zero as the problem size grows. In addition, \textsc{Greedy} typically runs within $1\%$ of the time of the MILP.

\section{Experiment Parameters}\label{app:params}
Table~\ref{tab:distribution} shows the distribution from which the parameters are generated in most of our experiments. In Table~\ref{tab:distribution2}, we detail the parameters used in the experiment in Fig.~\ref{fig:preprocess}.

\begin{table}[h]
\begin{center}
    \begin{tabular}{l  l}
    \hline
    Variable & Distribution \\ \hline
    $t^{all}_w$ & $U(350, 750)$\\ 
    $t_w$ & $U(50, 100)$\\
    $c_w$ &  $U(1,4)$\\
    $\pi_w$ & $U(30, 54)$\\
    $B_d$ & $U(0.11\sum_{w \in W} c_w t_w, 0.71\sum_{w \in W} c_w t_w)$\\
    $B_a$ & $U(0.1\sum_{w \in W} \pi_w, 0.8\sum_{w \in W} \pi_w)$ \\
    $B_e$ & $U(0.2\sum_{w \in W} t_w^{all}, 0.8\sum_{w \in W} t_w^{all})$\\ \hline
    \end{tabular}
\end{center}
\caption{Parameter distribution} \label{tab:distribution}
\end{table}

\begin{table}[h!]
    \centering
    \begin{tabular}{cccc}
        \cline{1-4}
         \multicolumn{2}{c}{For $w \in W_1$} & \multicolumn{2}{c}{For $w \in W_2$}\\
        \cline{1-4}
        Variable & Distribution & Variable & Distribution\\
        \cline{1-4}
         $t^{all}_w$ & $U(60, 110)$ & $t^{all}_w$ & $U(20, 70)$\\
         $t_w$ & $U(45, 55)$ & $t_w$ & $U(3, 8)$\\
         $c_w$ &  $U(2,6)$ & $c_w$ &  $U(1,3)$\\
         $\pi_w$ & $3$ & $\pi_w$ & $3$ \\
         \cline{1-4}
        $B_d$ &\multicolumn{3}{c} {$U(0, 10\sum_{w \in W} c_w t_w/|W|)$}\\
        $B_a$ &\multicolumn{3}{c} {$U(0.1\sum_{w \in W} \pi_w, 0.8\sum_{w \in W} \pi_w)$}\\
        $B_e$ &\multicolumn{3}{c} {$U(0, 3\sum_{w \in W} t_w^{all}/|W|)$}\\
        \hline
    \end{tabular}
    \caption{Parameter distributions for the experiment on large instances.} \label{tab:distribution2}
\end{table}

In addition, in the case of small effort budget, $B_e$ is generated uniformly between $1$ and $\min_{w \in W} t^w_{all}$. 

For large scale instances, we set different websites to have different importance, motivated by the fact that people do not visit all websites with equal frequency.
We split $W$ into $W_1, W_2$ with $|W_1|:|W_2| = 1:9$.
Websites in $W_1$ have a large portion of traffic from the organization and those in $W_2$ have a smaller portion. Thus, $W_1$ and $W_2$ follow different distributions (Table~\ref{tab:distribution2}). 
The attacker has a uniform cost of attack. In less than 4 of the 20 instances DWE did not reduce the problem size by much. We report in Fig.~\ref{fig:preprocess} the majority group where DWE eliminated a significant number of websites.
$|W_1|/|W|$ could be a lot smaller in reality, and our algorithms with DWE would run even faster.

\section{Discussion} \label{app:discussion}
\textbf{Assumptions and generality\,\,}
We assumed that the attack will succeed if and only if the network packet is unaltered. If the attacker can obtain the true system information with probability $p_w$ even if the packet is altered, we may modify the objective in Eq.~\eqref{lp1Eq1} to $\sum_{w} t_w(1- x_w(1-p_w))e_w/t^{all}_w$.
If the organization has other countermeasures (e.g. Bromium browser VMs), the attack may fail with probability $q_w$ even if the packet is unaltered, the objective then becomes $\sum_{w} t_w(1- x_w)(1-q_w)e_w/t^{all}_w$.
Thus, our algorithm can account for different levels of adversary and defender sophistication.

We do not attempt to claim that altering the network packets is a panacea to all watering hole attacks. 
Cyber attackers have many tools to circumvent existing deception techniques. 
Nonetheless, the proposed deception technique increases their uncertainty about the true nature of the environment, which leads to more cost on them, e.g.\ technical complexity and increased exposure.
This uncertainty ties into our consideration of the attacker's scanning effort $e_w$ and budget $B_e$, as the attacker cannot easily obtain or trust the basic information in the network packets.

\textbf{Limitations\,\,}
The generality notwithstanding, 
We acknowledge a few limitations of our work and potential problems in large-scale deployment. 
First, if an organization is the sole user of our method and if the attacker has (possibly imperfect) clue about the source of traffic from the start, randomizing network packet information might serve as an unintended signal to the attacker, reducing the effort needed $e_w$ to identify traffic from the targeted organization.
Second, by manipulating the web traffic, the organization is effectively monitoring its employees' internet activities. Although in many jurisdictions this is allowed when doing properly, the potential ethical issues must be carefully addressed.

\end{document}